\documentclass[journal]{IEEEtran}
\IEEEoverridecommandlockouts
\usepackage{lineno}
\usepackage{hyperref}
\usepackage{cite}
\usepackage{amsmath,amssymb,amsfonts,cases} 
\usepackage{amsmath}
\usepackage{amsthm}
\DeclareMathOperator*{\argmax}{argmax}

\usepackage{graphicx}
\usepackage{textcomp}
\usepackage{xcolor}
\usepackage{graphicx}
\usepackage{float}
\usepackage{subfigure}
\usepackage{amsmath,mleftright}
\usepackage{amsfonts,amssymb}
\usepackage{mathrsfs}
\usepackage{mathtools}
\usepackage{algorithm}
\usepackage{algorithmic}
\usepackage{bm}
\usepackage{multirow}
\usepackage{array}
\usepackage{amssymb}
\usepackage{amsmath}
\usepackage{cite}
\usepackage{url}
\usepackage{xcolor}
\usepackage{cite,graphicx,amsmath,amssymb}
\usepackage{subfigure}
\usepackage{fancyhdr}
\usepackage{mdwmath}
\usepackage{mdwtab}
\usepackage{caption}
\usepackage{amsthm}
\usepackage{setspace}
\usepackage{bm}
\usepackage{mathtools}
\usepackage{dsfont}
\usepackage{bbm}
\usepackage{framed}

\newtheorem{theorem}{Theorem}

\newtheorem{lemma}{Lemma}

\newtheorem{corollary}{Corollary}

\allowdisplaybreaks[4]
\makeatletter
\newcommand{\biggg}{\bBigg@{3}}
\newcommand{\Biggg}{\bBigg@{3.5}}
\makeatother
\makeatletter
\renewcommand{\maketag@@@}[1]{\hbox{\m@th\normalsize\normalfont#1}}%
\makeatother
\def\BibTeX{{\rm B\kern-.05em{\sc i\kern-.025em b}\kern-.08em
    T\kern-.1667em\lower.7ex\hbox{E}\kern-.125emX}}
    \expandafter\def\expandafter\normalsize\expandafter{%
    \normalsize%
    \setlength\abovedisplayskip{4pt}%
    \setlength\belowdisplayskip{4pt}%
    \setlength\abovedisplayshortskip{2pt}%
    \setlength\belowdisplayshortskip{2pt}%
}
\begin{document}
\title{Secure Multiuser Beamforming With Movable Antenna Arrays}
\author{Zhenqiao Cheng, Boqun Zhao, Chongjun Ouyang, and Xingqi Zhang\vspace{-10pt}
\thanks{Z. Cheng is with the 6G Research Centre, China Telecom Beijing Research Institute, Beijing 102209, China (e-mail: zhenqiao.cheng@engineer.com).}
\thanks{B. Zhao and X. Zhang are with Department of Electrical and Computer Engineering, University of Alberta, Edmonton AB, T6G 2R3, Canada (email: \{boqun1, xingqi.zhang\}@ualberta.ca).}
\thanks{C. Ouyang is with the School of Electronic Engineering and Computer Science, Queen Mary University of London, London, E1 4NS, U.K. (e-mail: c.ouyang@qmul.ac.uk).}}
\maketitle
\begin{abstract}
A movable-antenna (MA)-enabled secure multiuser transmission framework is developed to enhance physical-layer security. Novel expressions are derived to characterize the achievable sum secrecy rate based on the secure channel coding theorem. On this basis, a joint optimization algorithm for digital beamforming and MA placement is proposed to maximize the sum secrecy rate via fractional programming and block coordinate descent. In each iteration, every variable admits either a closed-form update or a low-complexity one-dimensional or bisection search, which yields an efficient implementation. Numerical results demonstrate the effectiveness of the proposed method and show that the MA-enabled design achieves higher secrecy rates than conventional fixed-position antenna arrays.
\end{abstract}
\section{Introduction}
\begin{figure}[!t]
\centering
\includegraphics[height=0.10\textwidth]{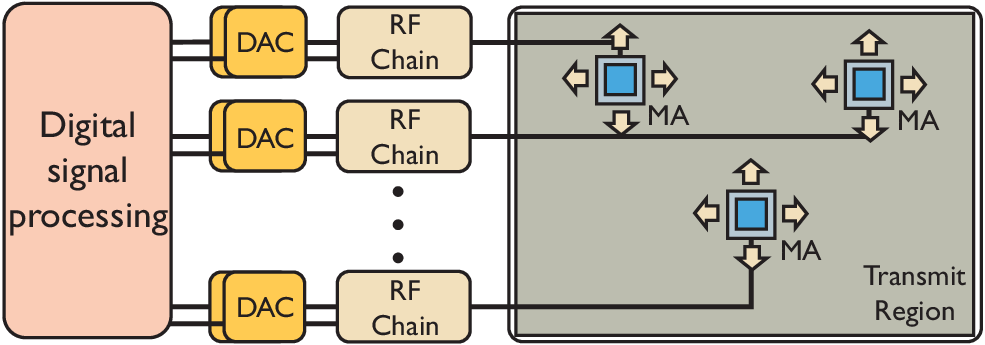}
\caption{Illustration of MA-empowered MIMO architecture.}
\label{Figure0}
\vspace{-10pt}
\end{figure}

Reconfigurable antennas have attracted increasing attention in recent years because they enable flexible control of the electromagnetic (EM) radiation characteristics of an antenna array. Such control can reshape the wireless channel and improve communication performance. A representative example is the movable antenna (MA) architecture, which adjusts antennas' physical positions to exploit spatial diversity and reduce the impact of small-scale fading, as illustrated in {\figurename} {\ref{Figure0}}. In recent years, MA-enabled system design and performance analysis have advanced rapidly, and existing results have demonstrated substantial throughput gains; see the recent survey for a detailed review \cite{zhu2025tutorial}. These developments suggest that MAs are a promising component for next-generation multiple-input multiple-output (MIMO) systems \cite{heath2025tri}.

Among existing studies on MAs, a widely investigated topic is physical-layer security enhancement \cite{cheng2024enabling}. With properly optimized MA locations, the radiated energy can be steered more effectively toward the legitimate users, and the information leakage to eavesdroppers can be reduced. Further gains can be achieved through joint design of the digital beamforming and the MA-position-induced electromagnetic beamforming, which improves the achievable secrecy rate \cite{cheng2024enabling,tang2025deep,le2025beamforming,wang2025movable,xiong2025secure,feng2024movable}. 

Early works have explored the secrecy benefits brought by MA deployment. However, most existing results focus on a single-user setting with one legitimate user and one eavesdropper \cite{cheng2024enabling}. Several studies have extended the analysis to multiuser scenarios, but they typically impose restrictive assumptions. Examples include single-stream multicast transmission \cite{xiong2025secure}, the single-eavesdropper case \cite{le2025beamforming}, and special channel conditions \cite{feng2024movable}. A general MA-enabled multiuser secure transmission framework that accommodates arbitrary numbers of legitimate users/data streams and cooperating eavesdroppers remains unavailable. The corresponding joint beamforming and MA placement design problem also remains open.

\begin{figure}[!t]
\centering
\includegraphics[height=0.15\textwidth]{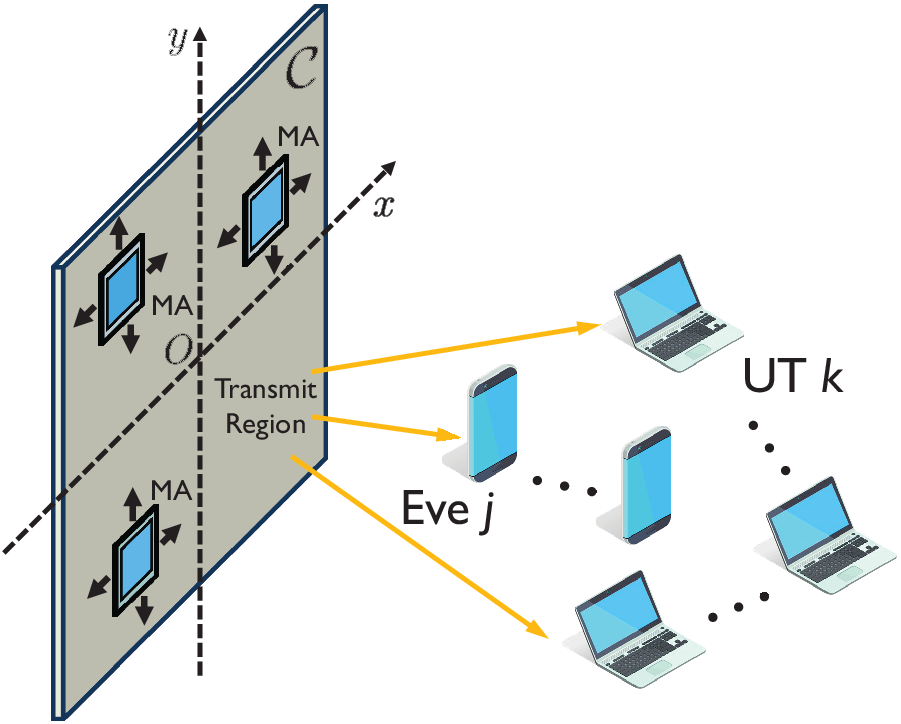}
\caption{Illustration of a multiuser wiretap channel.}
\label{Figure1}
\vspace{-15pt}
\end{figure}

To address this gap and deepen the understanding of MA-enabled secure communications, this article establishes an MA-enabled multiuser secure transmission framework, as illustrated in {\figurename} {\ref{Figure1}}. Within this framework, we characterize the achievable sum secrecy rate based on secure channel coding theory, and we develop a joint digital beamforming and MA placement algorithm that maximizes the sum secrecy rate through fractional programming (FP) \cite{shen2018fractional}. Numerical results verify the effectiveness of the proposed method and demonstrate clear secrecy-rate gains over conventional fixed-position antenna arrays (FPAs) in multiuser settings.

\section{System Model}
Consider an MA-empowered multiuser secure transmission scenario in which a multi-antenna base station (BS) simultaneously serves $K$ legitimate user terminals (UTs) over the same time-frequency resources. The transmitted messages must remain confidential against $J$ eavesdroppers (Eves), as shown in {\figurename} {\ref{Figure1}}. The BS employs $M$ transmit MAs, while each UT $k\in[K]\triangleq\{1,\ldots,K\}$ and each Eve $j\in[J]\triangleq\{1,\ldots,J\}$ is equipped with a single receive antenna. Each MA is connected to a dedicated radio-frequency (RF) chain through flexible cables, so its position can be adjusted in real time within a prescribed region, as illustrated in {\figurename} {\ref{Figure0}}. Let the position of the $m$th MA be ${\mathbf{t}}_m\triangleq[t_{m,x}, t_{m,y}]^{\mathsf{T}}\in{\mathcal{C}}\subseteq{\mathbbmss{R}}^{2\times1}$ for $m\in[M]\triangleq\{1,\ldots,M\}$, where $\mathcal{C}$ denotes the two-dimensional transmit region in which the MAs are allowed to move. Without loss of generality, $\mathcal{C}$ is a square area with size $A\times A$.
\subsection{Channel Model}
We assume quasi-static block fading and focus on a single fading block. Within this block, the multipath components remain unchanged for all locations in $\mathcal{C}$. Let ${\mathbf h}_k\in{\mathbbmss C}^{M\times1}$ and ${\mathbf g}_j\in{\mathbbmss C}^{M\times1}$ denote the BS channels associated with UT $k$ and Eve $j$, respectively. We adopt a field-response channel model and write ${\mathbf h}_{k}\triangleq[h_{k}({\mathbf{t}}_1),\ldots, h_{k}({\mathbf{t}}_M)]^{{\mathsf{T}}}$ and ${\mathbf g}_j\triangleq[g_{j}({\mathbf{t}}_1),\ldots , g_{j}({\mathbf{t}}_M)]^{{\mathsf{T}}}$, where
\begin{subequations}\label{Channel_Model}
\begin{align}
h_{k}({\mathbf{t}})&\triangleq\beta_{k,0}{\rm{e}}^{-{\rm{j}}\frac{2\pi}{\lambda}{\mathbf{t}}^{\mathsf{T}}{\bm\rho}_{k,0}}+\sum\nolimits_{\ell=1}^{L_k}\beta_{k,\ell}{\rm{e}}^{-{\rm{j}}\frac{2\pi}{\lambda}{\mathbf{t}}^{\mathsf{T}}{\bm\rho}_{k,\ell}},\\
g_{j}({\mathbf{t}})&\triangleq\hat{\beta}_{j,0}{\rm{e}}^{-{\rm{j}}\frac{2\pi}{\lambda}{\mathbf{t}}^{\mathsf{T}}\hat{\bm\rho}_{j,0}}+\sum\nolimits_{\ell=1}^{\hat{L}_j}\hat{\beta}_{j,\ell}{\rm{e}}^{-{\rm{j}}\frac{2\pi}{\lambda}{\mathbf{t}}^{\mathsf{T}}\hat{\bm\rho}_{j,\ell}}.
\label{Channel_Model2}
\end{align}
\end{subequations}
Here, $\beta_{k,0}$ and $\hat{\beta}_{j,0}$ denote the complex gains of the \emph{line-of-sight (LoS)} components, while $\beta_{k,\ell}$ and $\hat{\beta}_{j,\ell}$ denote the gains of the $\ell$th \emph{non-line-of-sight (NLoS)} paths. The spatial direction vectors are ${\bm\rho}_{k,\ell}\triangleq[\sin{\theta_{k,\ell}}\cos{\phi_{k,\ell}},\cos{\theta_{k,\ell}}]^{{\mathsf{T}}}$ and $\hat{\bm\rho}_{j,\ell}\triangleq[\sin{\hat{\theta}_{j,\ell}}\cos{\hat{\phi}_{j,\ell}},\cos{\hat{\theta}_{j,\ell}}]^{{\mathsf{T}}}$, where $\theta_{k,\ell}$ and $\hat{\theta}_{j,\ell}$ are the elevation angles, $\phi_{k,\ell}$ and $\hat{\phi}_{j,\ell}$ are the azimuth angles, and $L_k$ and $\hat{L}_j$ are the numbers of propagation paths.

The system operates in time-division duplexing (TDD) mode. The BS acquires instantaneous channel state information (CSI) through uplink pilot training. We consider passive eavesdropping in which the eavesdroppers are \emph{registered UTs but are not trusted by the legitimate users}. Under this assumption, the eavesdroppers transmit pilots during the training phase, and the BS estimates their CSI as well. We assume mutually orthogonal pilots and negligible estimation error, so the BS has perfect CSI for all UTs and Eves. The impact of imperfect Eve-CSI and the corresponding robust MA designs are left for future investigation.
\subsection{Multiuser Secure Transmission}
The BS aims to transmit a confidential message ${\mathsf{m}}_k\in[2^{T{\mathcal{R}}_k}]\triangleq\{1,\ldots,2^{T{\mathcal{R}}_k}\}$ to each UT $k\in[K]$ within $T$ channel uses, such that it is kept secret from the eavesdroppers. To this end, the BS first securely encodes confidential message ${\mathsf{m}}_k$ into the codeword $[s_k(1),\ldots,s_k(T)]$ using the encoder $f_{k,T}(\cdot):[2^{T{\mathcal{R}}_k}] \rightarrow {\mathbbmss{C}}^{T}$. The BS then maps the vector of encoded symbols in the time interval $t \in[T]\triangleq\{1,\ldots, T\}$, i.e., ${\mathbf{s}}(t) \triangleq [s_1(t),\ldots,s_K(t)]^{\mathsf{T}}$, into the transmit signal ${\mathbf{x}}(t)\in{\mathbbmss{C}}^{M\times1}$ which is sent towards the UTs while being overheard by Eves. Let ${\mathbf{w}}_k\in{\mathbbmss{C}}^{M\times1}$ denote the beamforming vector associated with UT $k$. The above arguments imply that ${\mathbf{x}}(t) = {\mathbf{W}}{\mathbf{s}}(t) = \sum_{k=1}^{K}{\mathbf{w}}_ks_k(t)$, where ${\mathbf{W}} \triangleq [{\mathbf{w}}_1, \ldots , {\mathbf{w}}_K]\in{\mathbbmss{C}}^{M\times K}$. As a result, the observations at UT $k$ and Eve $j$ are, respectively, given by
\begin{align}
&y_k(t)={\mathbf h}_k^{\mathsf T}{\mathbf{x}}(t)+n_k(t)={\mathbf h}_k^{\mathsf T}\sum\nolimits_{i=1}^{K}\mathbf{w}_is_i(t)+n_k(t),\\
&\hat{y}_j(t)={\mathbf g}_j^{\mathsf T}{\mathbf{x}}(t)+\hat{n}_j(t)={\mathbf g}_j^{\mathsf T}\sum\nolimits_{i=1}^{K}\mathbf{w}_is_i(t)+\hat{n}_j(t),
\end{align}
where $n_k(t)\sim{\mathcal{CN}}(0,\sigma_k^2)$ and $\hat{n}_j(t)\sim{\mathcal{CN}}(0,\hat{\sigma}_j^2)$ denote the $t$th sample of the additive noises for $ t \in[T]$ with $\sigma_k^2$ and $\hat{\sigma}_j^2$ being the noise powers. 

Assume that each UT $k$ and each Eve $j$ can obtain instantaneous CSI of its own channel through properly designed pilot signals. Each UT $k\in[K]$ employs a decoder $\hat{f}_{k,T}(\cdot):{\mathbbmss{C}}^{T}\rightarrow[2^{T{\mathcal{R}}_k}]$ and produces an estimate $\hat{\mathsf{m}}_k$ of the transmitted message ${\mathsf{m}}_k$ from the received samples $[y_k(1),\ldots,y_k(T)]$. The corresponding block error probability is ${\mathcal{P}}_k \triangleq \Pr(\hat{\mathsf{m}}_k \ne {\mathsf{m}}_k)$. The confidential message ${\mathsf{m}}_k$ is also observed through the eavesdroppers' channels. For a worst-case design, we assume that the $J$ eavesdroppers cooperate to intercept the transmission. This corresponds to a worst-case secrecy scenario, while non-cooperative eavesdroppers would generally result in higher achievable secrecy rates. We further assume that they can remove multiuser interference from the other UTs. After observing $[\hat{\mathbf{y}}_1,\ldots,\hat{\mathbf{y}}_J]$ with $\hat{\mathbf{y}}_j\triangleq[\hat{y}_j(1),\ldots,\hat{y}_j(T)]$ for all $j\in[J]$, the eavesdroppers retain an average residual uncertainty $H({\mathsf{m}}_k|[\hat{\mathbf{y}}_1,\ldots,\hat{\mathbf{y}}_J])$. The encoded symbols are independent and identically distributed (i.i.d.) with zero mean and unit variance. The resulting signal-to-interference-plus-noise ratio (SINR) at UT $k$ is given by $\gamma_k=\frac{\lvert{\mathbf h}_k^{\mathsf T}{\mathbf w}_k\rvert^2}{\sigma_k^2+\sum_{k'\neq k}\lvert{\mathbf h}_k^{\mathsf T}{\mathbf w}_{k'}\rvert^2}$. The aggregated SNR at the cooperating eavesdroppers for decoding ${\mathsf{m}}_k$ is given by ${\hat{\gamma}}_k=\sum\nolimits_{j=1}^{J}\frac{1}{\hat{\sigma}_j^2}\lvert{\mathbf g}_j^{\mathsf T}{\mathbf w}_k\rvert^2$.

A perfect secrecy rate ${\mathcal{R}}_k$ for UT $k$ is achievable if there exists an encoder-decoder pair such that, as $T \rightarrow\infty$, the decoding error probability satisfies ${\mathcal{P}}_k\rightarrow0$ and the equivocation satisfies $\Delta_k\triangleq\frac{H({\mathsf{m}}_k|[\hat{\mathbf{y}}_1,\ldots,\hat{\mathbf{y}}_J])}{H({\mathsf{m}}_k)}=\frac{H({\mathsf{m}}_k|[\hat{\mathbf{y}}_1,\ldots,\hat{\mathbf{y}}_J])}{T{\mathcal{R}}_k}\rightarrow1$. The condition $\Delta_k\rightarrow1$ is equivalent to $H({\mathsf{m}}_k|[\hat{\mathbf{y}}_1,\ldots,\hat{\mathbf{y}}_J]) \rightarrow H({\mathsf{m}}_k)$. It also implies $I({\mathsf{m}}_k;[\hat{\mathbf{y}}_1,\ldots,\hat{\mathbf{y}}_J]) = H({\mathsf{m}}_k)-H({\mathsf{m}}_k|[\hat{\mathbf{y}}_1,\ldots,\hat{\mathbf{y}}_J]) \rightarrow0$, so the eavesdroppers cannot obtain information about ${\mathsf{m}}_k$. By the secure coding theorem, an achievable secrecy rate can be obtained via Gaussian random coding. In particular, the transmitted symbols are i.i.d. with ${s}_k(t)\sim{\mathcal{CN}}(0,1)$ for $t\in[T]$ and $k\in[K]$. For a given beamforming matrix $\mathbf{W}$, channel set $\{\mathbf{h}_k\}$, and eavesdropper channel set $\{\mathbf{g}_j\}$, the secrecy rate of UT $k$ is given by \cite{leung2003gaussian}
\begin{align}\label{Secrecy_Rate_UT}
{\mathcal R}_{k}=[\log_2\left(1+\gamma_k\right)-\log_2\left(1+{\hat{\gamma}}_k\right),0]^{+},
\end{align}
where $[\cdot]^{+}\triangleq\max\{\cdot,0\}$. This expression characterizes the largest confidential transmission rate that supports reliable decoding at UT $k$ while ensuring perfect secrecy against the cooperating eavesdroppers. With superposition coding across UTs, the achievable sum secrecy rate is given as follows:
\begin{align}
{\mathcal R}=\sum\nolimits_{k=1}^{K}[\log_2\left(1+\gamma_k\right)-\log_2\left(1+{\hat{\gamma}}_k\right),0]^{+}\label{Secrecy_Rate_System}.
\end{align}
It is important to note that, unlike in FPA-based wiretap channels, the secrecy rate in MA-based channels depends on the physical positions of the MAs.
\subsection{Problem Formulation}
Based on existing MA channel-estimation methods \cite{ma2023compressed}, the elevation and azimuth angles $\{\theta_{k,\ell},\hat{\theta}_{j,\ell},\phi_{k,\ell},\hat{\phi}_{j,\ell}\}$, as well as the path gains $\{\beta_{k,\ell},\hat{\beta}_{j,\ell}\}$, in \eqref{Channel_Model}, can be accurately estimated. With this channel knowledge, we jointly optimize the transmit beamforming matrix ${\mathbf{W}}$ and the MA locations ${\mathbf{T}}\triangleq[{\mathbf{t}}_1,\ldots,{\mathbf{t}}_M]\in{\mathbbmss{R}}^{2\times M}$ to maximize the sum secrecy rate in \eqref{Secrecy_Rate_System}. The resulting problem is formulated as follows:
\begin{subequations}\label{P_1}
\begin{align}
&\max\nolimits_{{\mathbf{T}},{}\mathbf{W}}~{\mathcal R}\\
&~{\rm{s.t.}}~{\mathsf{tr}}({\mathbf W}{\mathbf W}^{\mathsf H})=\sum\nolimits_{k=1}^{K}{\mathbf w}_k^{\mathsf H}{\mathbf w}_k\leq p,\label{P_1_c1}\\
&~\quad~~{\mathbf{t}}_m\in{\mathcal{C}},\forall m\in[M],\lVert {\mathbf{t}}_m-{\mathbf{t}}_{m'}\rVert\geq D,\forall m\ne m',\label{P_1_c2}
\end{align}
\end{subequations}
where $p$ is the transmit power budget, and $D$ is the minimum antenna spacing that mitigates strong mutual coupling. Problem \eqref{P_1} is difficult because the objective $\mathcal{R}$ is non-concave in $\mathbf{T}$ and $\mathbf W$, and constraint \eqref{P_1_c2} is non-convex. The variables $\mathbf{T}$ and $\mathbf W$ are also strongly coupled through the channels in \eqref{Channel_Model}. We next develop an efficient framework that yields a high-quality feasible solution with moderate computational complexity.
\section{Secure Multiuser Beamforming Design}
By invoking the FP framework \cite{shen2018fractional}, we transform problem \eqref{P_1} into an equivalent form that is more tractable. We then solve the transformed problem via a block coordinate descent (BCD) method, which separates the optimization variables and mitigates the strong coupling between $\mathbf{T}$ and $\mathbf W$.
\subsection{Reformulation of Problem \eqref{P_1}}
The operator $[\cdot]^{+}$ in \eqref{Secrecy_Rate_System} makes the objective in \eqref{P_1} non-smooth and difficult to optimize. To remove this operator, we rewrite \eqref{P_1} in an equivalent form as follows.
\vspace{-5pt}
\begin{lemma}\label{Lemma_Initial_Trans}
Let ${\mathbf b}=[b_1,\ldots,b_K]^{\mathsf T}$. Problem \eqref{P_1} is equivalent to the problem defined as follows:
\begin{subequations}\label{P_2}
\begin{align}
\max_{{\mathbf W},{\mathbf T},{\mathbf b}}~&\hat{\mathcal R}=\sum\nolimits_{k=1}^{K}b_k(\log_2\left(1+\gamma_k\right)-\log_2\left(1+{\hat{\gamma}}_k\right))\\
\quad{\rm{s.t.}}~&\eqref{P_1_c1},\eqref{P_1_c2},b_k\in[0,1],\forall k\in[K].\label{P_2_C}
\end{align}
\end{subequations}
Moreover, the optimal $\{{\mathbf W},{\mathbf T}\}$ is identical for \eqref{P_1} and \eqref{P_2}.
\end{lemma}
\vspace{-5pt}
\begin{IEEEproof}
We establish the claim by showing that problem \eqref{P_2} reduces to problem \eqref{P_1} after optimizing $\mathbf{b}$. Given $\{{\mathbf W},{\mathbf{T}}\}$, the optimal $b_k$ satisfies $b_k^{\star}={\mathbf 1}_{\{\gamma_k>{\hat{\gamma}}_k\}}$ with ${\mathbf 1}_{\{\cdot\}}$ denoting the indicator function. Inserting $b_k=b_k^{\star}$ into $\hat{\mathcal R}$ gives
\begin{align}
\hat{\mathcal R}
&=\sum\nolimits_{k=1}^{K}{\mathbf 1}_{\{\gamma_k>{\hat{\gamma}}_k\}}[\log_2\left(1+\gamma_k\right)-\log_2\left(1+{\hat{\gamma}}_k\right)]\nonumber\\
&=\sum\nolimits_{k=1}^{K}[\log_2\left(1+\gamma_k\right)-\log_2\left(1+{\hat{\gamma}}_k\right),0]^{+}={\mathcal R},\nonumber
\end{align}
which is exactly the objective in \eqref{P_1}.
\end{IEEEproof}
We further rewrite the objective in \eqref{P_2} as follows:
\begin{align}
\hat{\mathcal {R}}=\sum\nolimits_{k=1}^{K}b_k\left(\log_2\left(1+\gamma_k\right)+\log_2\left(1/({1+{\hat{\gamma}}_k})\right)\right).
\end{align}
This sum-of-ratios structure motivates the use of FP to obtain a more tractable formulation. The following result is used in the subsequent derivations.
\vspace{-5pt}
\begin{lemma}\label{Lemma_FP_Must}
Define $g\triangleq\sum_{j=1}^{J}\frac{pM(\hat{L}_j+1)}{\hat{\sigma}_j^2}\sum_{\ell=0}^{\hat{L}_j}\lvert\hat{\beta}_{j,\ell}\rvert^2$. Problem \eqref{P_2} is equivalent to the problem defined as follows:
\begin{subequations}\label{P_3}
\begin{align}
\max_{{\mathbf W},{\mathbf T},{\mathbf b}}&\sum_{k=1}^{K}b_k\left(\log_2\left(1+\gamma_k\right)+\log_2\left(1+\frac{g-{\hat{\gamma}}_k}{1+{\hat{\gamma}}_k}\right)\right)\\
\quad{\rm{s.t.}}~&\eqref{P_2_C},
\end{align}
\end{subequations}
where $g-{\hat{\gamma}}_k\geq0$ for $k\in[K]$.
\end{lemma}
\vspace{-5pt}
\begin{IEEEproof}
From the Cauchy-Schwarz inequality, we have 
\begin{align}\nonumber
{\hat{\gamma}}_k=\sum\nolimits_{j=1}^{J}\lvert{\mathbf g}_j^{\mathsf T}{\mathbf w}_k\rvert^2/{\hat{\sigma}_j^2}\leq
\sum\nolimits_{j=1}^{J}\lVert{\mathbf g}_j\rVert^2\lVert{\mathbf w}_k\rVert^2/{\hat{\sigma}_j^2}.
\end{align}
Under the power constraint \eqref{P_1_c1}, $\lVert{\mathbf w}_k\rVert^2\leq\sum\nolimits_{i=1}^{K}{\mathbf w}_i^{\mathsf H}{\mathbf w}_k\leq p$. Therefore, it follows that ${\hat{\gamma}}_k\leq
\sum\nolimits_{j=1}^{J}\frac{1}{\hat{\sigma}_j^2}\lVert{\mathbf g}_j\rVert^2\lVert{\mathbf w}_k\rVert^2\leq
\sum\nolimits_{j=1}^{J}\frac{p}{\hat{\sigma}_j^2}\lVert{\mathbf g}_j\rVert^2$. Next, it follows from \eqref{Channel_Model2} that, for any MA position ${\mathbf{t}}_m$, the channel magnitude satisfies $\lvert g_{j}({\mathbf{t}}_m)\rvert^2=\lvert\sum\nolimits_{\ell=0}^{\hat{L}_j}\hat{\beta}_{j,\ell}{\rm{e}}^{-{\rm{j}}\frac{2\pi}{\lambda}{\mathbf{t}}^{\mathsf{T}}\hat{\bm\rho}_{j,\ell}}\rvert^2\leq(\hat{L}_j+1)\sum\nolimits_{\ell=0}^{\hat{L}_j}\lvert\hat{\beta}_{j,\ell}\rvert^2$, which yields
\begin{align}\nonumber
\lVert{\mathbf g}_j\rVert^2=\sum\nolimits_{m=1}^{M}\lvert g_{j}({\mathbf{t}}_m)\rvert^2\leq M(\hat{L}_j+1)\sum\nolimits_{\ell=0}^{\hat{L}_j}\lvert\hat{\beta}_{j,\ell}\rvert^2.
\end{align}
It follows that ${\hat{\gamma}}_k\leq\sum\nolimits_{j=1}^{J}\frac{p}{\hat{\sigma}_j^2}M(\hat{L}_j+1)\sum\nolimits_{\ell=0}^{\hat{L}_j}\lvert\hat{\beta}_{j,\ell}\rvert^2=g$. Since $\log_2\left(1+g\right)$ does not depend on $\{{\mathbf W},{\mathbf T}\}$, this bound enables the equivalent reformulation from \eqref{P_2} to \eqref{P_3}.
\end{IEEEproof}
We next apply the FP framework to obtain an equivalent formulation of \eqref{P_3} with a more tractable objective.
\vspace{-5pt}
\begin{lemma}\label{Lemma_Dual}
Problem \eqref{P_3} is equivalent to:
\begin{subequations}\label{P_4}
\begin{align}
\max_{{\mathbf W},{\mathbf T},{\mathbf b},{\bm\alpha},{\bm\beta}}&
{\mathcal F}({\mathbf W},{\mathbf T},{\mathbf b},{\bm\alpha},{\bm\beta})\triangleq
\sum\nolimits_{k=1}^{K}b_k(f_{1}^{k}+f_{2}^{k})\\
\quad{\rm{s.t.}}~&\eqref{P_2_C},
\end{align}
\end{subequations}
where ${\bm\alpha}\triangleq[\alpha_1,\ldots,\alpha_K]^{\mathsf T}$, ${\bm\beta}\triangleq[\beta_1,\ldots,\beta_K]^{\mathsf T}$,
\begin{align}
&f_{1}^{k}\triangleq\log(1+\alpha_k)-\alpha_k+\frac{(1+\alpha_k)\lvert{\mathbf h}_k^{\mathsf T}{\mathbf w}_k\rvert^2}{\sum_{i=1}^{K}\lvert{\mathbf h}_k^{\mathsf T}{\mathbf w}_{i}\rvert^2+\sigma_k^2},\\
&f_{2}^{k}\triangleq\log(1+\beta_k)-\beta_k+{(1+\beta_k)(g-{\hat{\gamma}}_k)}
({1+g})^{-1}.
\end{align}
The optimal auxiliary variables satisfy $\alpha_k^{\star}=\gamma_k$ and $\beta_k^{\star}=\frac{g-{\hat{\gamma}}_k}{1+{\hat{\gamma}}_k}$ for $k\in[K]$.
\end{lemma}
\vspace{-5pt}
\begin{IEEEproof}
For fixed ${\mathbf W}$, ${\mathbf T}$, and ${\mathbf b}$, the function ${\mathcal F}(\cdot)$ is concave in each scalar variable $\alpha_k$ and $\beta_k$. The optimal $\alpha_k$ and $\beta_k$ follow from the first-order optimality conditions $\frac{\partial}{\partial \alpha_k}{\mathcal F}=0$ and $\frac{\partial}{\partial \beta_k}{\mathcal F}=0$, which are easily seen as $\alpha_k^{\star}$ and $\beta_k^{\star}$. Inserting $\alpha_k^{\star}$ and $\beta_k^{\star}$ back to $\mathcal F$ recovers the objective function in \eqref{P_3}, thus establishing the equivalence of these two problems.
\end{IEEEproof}
Since $g$ does not depend on $\{{\mathbf W},{\mathbf T},{\mathbf b}\}$, the main difficulty in maximizing ${\mathcal F}(\cdot)$ comes from the fractional structure in $\{f_{1}^{k}\}_{k=1}^{K}$. We introduce an additional set of auxiliary variables ${\bm\eta}\triangleq[\eta_1,\ldots,\eta_K]^{\mathsf T}$ to further simplify the objective.
\vspace{-5pt}
\begin{lemma}\label{Lemma_Quad}
Problem \eqref{P_4} is equivalent to the following:
\begin{align}\label{P_5}
\max_{{\mathbf W},{\mathbf T},{\mathbf b},{\bm\alpha},{\bm\beta},{\bm\eta}}
\sum\nolimits_{k=1}^{K}b_k(g_{1}^{k}+f_{2}^{k})\quad{\rm{s.t.}}~\eqref{P_2_C},
\end{align}
where $g_{1}^{k}\triangleq\log(1+\alpha_k)-\alpha_k+(1+\alpha_k)\hat{g}_{1}^{k}$ with $\hat{g}_{1}^{k}\triangleq2\Re\{{\eta}_k^{*}{\mathbf h}_k^{\mathsf T}{\mathbf w}_k\}
-\lvert\eta_k\rvert^2(\sum\nolimits_{i=1}^{K}\lvert{\mathbf h}_k^{\mathsf T}{\mathbf w}_{i}\rvert^2+\sigma_k^2)$. The optimal $\eta_k$ is given by $\eta_k^{\star}=\frac{{\mathbf h}_k^{\mathsf T}{\mathbf w}_k}{\sum_{i=1}^{K}\lvert{\mathbf h}_k^{\mathsf T}{\mathbf w}_{i}\rvert^2+\sigma_k^2}$.
\end{lemma}
\vspace{-5pt}
\begin{IEEEproof}
The proof follows the same steps as Lemma \ref{Lemma_Dual}.
\end{IEEEproof}
Lemmas \ref{Lemma_Initial_Trans} to \ref{Lemma_Quad} show that \eqref{P_5} provides a variational representation of the original problem \eqref{P_1}. In particular, the optimal $\{{\mathbf W},{\mathbf T}\}$ obtained from \eqref{P_5} is also optimal for \eqref{P_1}.
\subsection{The Proposed BCD-Based Algorithm}
In problem \eqref{P_5}, the optimization variables are strongly coupled. This structure motivates the use of BCD, where each variable block is updated while the remaining blocks are fixed. We next describe the update for each block.
\subsubsection{Subproblem in $\mathbf W$}
For fixed ${\mathbf{T}}$, ${\mathbf b}$, ${\bm\alpha}$, ${\bm\beta}$, and ${\bm\eta}$, problem \eqref{P_5} reduces to the following:
\begin{align}\label{P_8}
\min_{\mathbf W}~
\sum\nolimits_{k=1}^{K}({\mathbf w}_k^{\mathsf H}{\mathbf A}_k{\mathbf w}_k-2\Re\{{\mathbf w}_k^{\mathsf H}{\mathbf a}_k\})\quad
{\rm{s.t.}}~\eqref{P_1_c1},
\end{align}
where ${\mathbf a}_k\triangleq b_k(1+\alpha_k){\eta}_k{\mathbf h}_k^{*}$ and ${\mathbf A}_k\triangleq\sum\nolimits_{i=1}^{K}b_i(1+\alpha_i)|\eta_i|^2{\mathbf h}_i^{*}{\mathbf h}_i^{\mathsf T}
+{b_k(1+\beta_k)}{({1+g})^{-1}}\sum\nolimits_{j=1}^{J}\frac{1}{\hat{\sigma}_j^2}{\mathbf g}_j^{*}{\mathbf g}_j^{\mathsf T}\succeq{\mathbf 0}$. Problem \eqref{P_8} is a convex quadratically constrained quadratic program (QCQP). The Karush-Kuhn-Tucker (KKT) conditions give the optimal beamformer ${\mathbf w}_k^{\star}=({\mathbf{A}}_k+\lambda{\mathbf I})^{-1}{\mathbf{a}}_k$ for $k\in[K]$, where $\lambda$ is the dual variable associated with the constraint ${\mathsf{tr}}({\mathbf W}{\mathbf W}^{\mathsf H})\leq p$. The complementary slackness condition satisfies $\lambda({\mathsf{tr}}({\mathbf{W}}{\mathbf{W}}^{\mathsf H})- p)=0$. If $\sum_{k=1}^{K}{\mathbf a}_k^{\mathsf H}{\mathbf A}_k^{-2}{\mathbf a}_k\leq p$, then $\lambda=0$. Otherwise, $\lambda>0$ and is chosen to meet the power constraint with equality, i.e., ${\mathsf{tr}}({\mathbf{W}}{\mathbf{W}}^{\mathsf H})=\sum\nolimits_{k=1}^{K}{\mathbf a}_k^{\mathsf H}({\mathbf{A}}_k+\lambda{\mathbf I})^{-2}{\mathbf a}_k=p$. Let ${\mathbf A}_k={\mathbf U}_{{\mathbf A}_k}{\bm\Lambda}_{{\mathbf A}_k}{\mathbf U}_{{\mathbf A}_k}^{\mathsf H}$ be the eigen-decomposition of ${\mathbf A}_k$, where ${\mathbf U}_{{\mathbf A}_k}{\mathbf U}_{{\mathbf A}_k}^{\mathsf H}={\mathbf{I}}$ and ${\bm\Lambda}_{{\mathbf A}_k}$ is diagonal with nonnegative entries $\{[{\bm\Lambda}_{{\mathbf A}_k}]_{m,m}\}_{m=1}^{M}$. Then, it follows that
\begin{align}
\sum\nolimits_{k=1}^{K}\sum\nolimits_{m=1}^{M}\frac{|[{\mathbf U}_{{\mathbf A}_k}^{\mathsf{H}}{\mathbf a}_k]_{m}|^2}{([{\bm\Lambda}_{{\mathbf A}_k}]_{m,m}+\lambda)^{2}}=p.\label{Power_Bisection}
\end{align}
The left-hand side of \eqref{Power_Bisection} is strictly decreasing in $\lambda\geq0$. Therefore, $\lambda$ can be found by a bisection search.

\begin{algorithm}[!t]
  \algsetup{linenosize=\tiny}
  \scriptsize
  \caption{Element-wise algorithm for solving problem \eqref{P_9i}}
  \label{Algorithm1}
  \begin{algorithmic}[1]
    \STATE initialize the optimization variables
    \REPEAT
    \FORALL{$m=1:M$} 
    \STATE update $t_m^x$ and $t_m^y$ through one-dimensional search
    \ENDFOR
    \UNTIL{convergence}
  \end{algorithmic}
\end{algorithm}

\subsubsection{Subproblem in $\mathbf T$}
With ${\mathbf W}$, ${\mathbf b}$, ${\bm\alpha}$, ${\bm\beta}$, and ${\bm\eta}$ fixed, the update of ${\mathbf{T}}=[{\mathbf{t}}_1,\ldots,{\mathbf{t}}_M]$ depends on ${\mathbf h}_{k}=[h_{k}({\mathbf{t}}_1),\ldots, h_{k}({\mathbf{t}}_M)]^{{\mathsf{T}}}$ and ${\mathbf g}_j=[g_{j}({\mathbf{t}}_1),\ldots , g_{j}({\mathbf{t}}_M)]^{{\mathsf{T}}}$ with ${\mathbf{t}}_m=[t_{m,x}, t_{m,y}]^{\mathsf{T}}$ for $m\in[M]$. Constants that do not involve $\mathbf{T}$ can be dropped, yielding the subproblem as follows:
\begin{subequations}\label{P_9i}
\begin{align}
\max_{{\mathbf T}}~&
\sum\nolimits_{k=1}^{K}b_k((1+\alpha_k)(2\Re\{{\eta}_k^{*}{\mathbf h}_k^{\mathsf T}{\mathbf w}_k\}-\lvert\eta_k\rvert^2\sum\nolimits_{i=1}^{K}\nonumber\\
&\times\left.\left.\lvert{\mathbf h}_k^{\mathsf T}{\mathbf w}_i\rvert^2\right)-{(1+\beta_k)}{(1+g)^{-1}}{\hat{\gamma}}_k\right)\\
{\rm{s.t.}}~&\eqref{P_1_c2}.
\end{align}
\end{subequations}
The coupling among the entries of $\mathbf{T}$ prevents a direct closed-form update. To mitigate this coupling, we use an element-wise alternating optimization strategy. Each coordinate $t_m^x$ and $t_m^y$ is updated while the other entries of $\mathbf{T}$ are fixed. For fixed $\mathbf{T}$ except $t_m^x$, the update of $t_m^x$ can be expressed as follows:
\begin{subequations}\label{P_9}
\begin{align}
\max_{{t_m^x}}~&
f_{m}^{x}(t_m^x)\triangleq\sum\nolimits_{k=1}^{K}b_k((1+\alpha_k)(2\Re\{{\eta}_k^{*}[{\mathbf w}_k]_mh_k^{x}(t_m^x)\}\nonumber\\
&-\left.\lvert\eta_k\rvert^2\sum\nolimits_{i=1}^{K}\lvert A_{i,k}^{m}+[{\mathbf w}_i]_mh_k^{x}(t_m^x)\rvert^2\right)\nonumber\\
&-\left.\frac{1+\beta_k}{1+g}\sum\nolimits_{j=1}^{J}\frac{1}{\hat{\sigma}_j^2}\lvert B_{k,j}^{m}+[{\mathbf w}_k]_mg_j^{x}(t_m^x)\rvert^2\right)\\
{\rm{s.t.}}~&t_m^x\in[-A/2,A/2],\\
&(t_m^x-t_{m'}^x)^2+(t_m^y-t_{m'}^y)^2\geq D^2,m\ne m',
\end{align}
\end{subequations}
where $h_k^{x}(t_m^x)\triangleq\sum\nolimits_{\ell=0}^{L_k}\beta_{k,\ell}{\rm{e}}^{-{\rm{j}}\frac{2\pi}{\lambda}(t_m^y\cos{\theta_{k,\ell}}+t_m^x\sin{\theta_{k,\ell}}\cos{\phi_{k,\ell}})}$, $g_j^{x}(t_m^x)\triangleq\sum\nolimits_{\ell=0}^{\hat{L}_j}\hat{\beta}_{j,\ell}{\rm{e}}^{-{\rm{j}}\frac{2\pi}{\lambda}(t_m^y\cos{\hat{\theta}_{k,\ell}}+t_m^x\sin{\hat{\theta}_{k,\ell}}\cos{\hat{\phi}_{k,\ell}})}$, $A_{i,k}^{m}\triangleq\sum_{m'\ne m}[{\mathbf w}_i]_{m'}h_k({\mathbf{t}}_{m'})$, and $B_{k,j}^{m}\triangleq\sum_{m'\ne m}[{\mathbf w}_k]_{m'}g_j({\mathbf{t}}_{m'})$. Problem \eqref{P_9} is non-convex. A near-optimal update can be obtained through a one-dimensional search over a finite grid. Let $Q$ denote the number of grid points and define ${\mathcal{Q}}\triangleq\{-\frac{A}{2},-\frac{A}{2}+\frac{A}{Q-1},\ldots,\frac{A}{2}\}$. The feasible search set for $t_m^x$ is given by
\begin{align}\nonumber
{\mathcal{Q}}_m^{x}\triangleq\{t\in{\mathcal{Q}}:(t-t_{m'}^x)^2+(t_m^y-t_{m'}^y)^2\geq D^2,\forall m\ne m'\}.
\end{align}
Then, a near-optimal solution for $t_m^x$ is given by $t_m^x=\argmax\nolimits_{t\in{\mathcal{Q}}_m^{x}}f_{m}^{x}(t)$. The update of $t_m^y$ follows the same steps. Algorithm \ref{Algorithm1} summarizes the resulting element-wise alternating optimization method for solving problem \eqref{P_9i}.

\subsubsection{Subproblem in ${\mathbf b}$}
With the remaining blocks fixed, the optimization over $\{b_k\}_{k=1}^{K}$ decouples into $K$ scalar problems. For each $k$, the update is given by
\begin{equation}\label{Problem_Weights}
\begin{split}
b_k^{\star}=\argmax\nolimits_{b_k\in[0,1]}(b_k(g_{1}^{k}+f_{2}^{k}))={\mathbf 1}_{\{g_1^{k}>-f_2^{k}\}}.
\end{split}
\end{equation}
\subsubsection{Subproblems in the Auxiliary Variables}
For given ${\mathbf W}$, ${\mathbf T}$, and ${\mathbf b}$, the auxiliary variables $\bm\alpha$, $\bm\beta$, and $\bm\eta$ admit the following closed-form updates from Lemmas \ref{Lemma_Dual} and \ref{Lemma_Quad}:
\begin{align}\label{Solution_Auxiliary} 
\alpha_k^{\star}=\gamma_k,~\beta_k^{\star}=\frac{g-{\hat{\gamma}}_k}{1+{\hat{\gamma}}_k},~\eta_k^{\star}=\frac{{\mathbf h}_k^{\mathsf T}{\mathbf w}_k}{\sum_{i=1}^{K}\lvert{\mathbf h}_k^{\mathsf T}{\mathbf w}_{i}\rvert^2+\sigma_k^2}.
\end{align}

\begin{algorithm}[!t]
  \algsetup{linenosize=\tiny}
  \scriptsize
  \caption{BCD-based algorithm for solving problem \eqref{P_1}}
  \label{Algorithm2}
  \begin{algorithmic}[1]
    \STATE initialize the optimization variables
    \REPEAT
    \STATE update ${\mathbf{W}}$ by KKT conditions
    \STATE update ${\mathbf{T}}$ by element-wise alternating optimization
    \STATE update $\mathbf{b}$, $\bm\alpha$, $\bm\beta$, and $\bm\eta$ by their closed-form solutions
    \UNTIL{convergence}
  \end{algorithmic}
\end{algorithm}

\subsubsection{Overall Algorithm, Convergence, and Complexity}
Combining the updates above yields the BCD procedure summarized in Algorithm \ref{Algorithm2}. The objective value is non-decreasing after each block update, and it is upper bounded. Therefore, the objective sequence generated by Algorithm \ref{Algorithm2} converges. The per-iteration computational complexities for updating ${\mathbf{W}}$, ${\mathbf{T}}$, $\mathbf{b}$, $\bm\alpha$, $\bm\beta$, and $\bm\eta$ are ${\mathcal{O}}(M^3+M^2(K+J))$, ${\mathcal{O}}(I_{\mathbf{T}}M\Xi_{\mathbf{T}})$, ${\mathcal{O}}(KM(K+J))$, ${\mathcal{O}}(K)$, ${\mathcal{O}}(K)$, and ${\mathcal{O}}(K)$, respectively, where $I_{\mathbf{T}}$ is the number of iterations used by Algorithm \ref{Algorithm1}, and $\Xi_{\mathbf{T}}$ denotes the cost of one full sweep of the element-wise search over $\mathbf{T}$. Under the grid search with $Q$ points and the field-response model in \eqref{Channel_Model}, $\Xi_{\mathbf{T}}$ scales on the order of ${\mathcal{O}}(KM(K+J)+Q(\sum_{k=1}^{K}L_k+\sum_{j=1}^{J}\hat{L}_j))$. 

\section{Numerical Results}
This section presents numerical results that verify the effectiveness of the proposed algorithm. For each UT $k$ and Eve $j$, the spatial channel includes one LoS component and $L_k=\hat{L}_j = 3$ NLoS components. The channel gains are generated as $\beta_{k,0}\sim{\mathcal{CN}}(0,1)$, $\hat{\beta}_{j,0}\sim{\mathcal{CN}}(0,1)$, $\beta_{k,\ell}\sim{\mathcal{CN}}(0,10^{-1})$ for $\ell=1,\ldots,L_k$, and $\hat{\beta}_{j,\ell}\sim{\mathcal{CN}}(0,10^{-1})$ for $\ell=1,\ldots,\hat{L}_j$. The elevation and azimuth angles $\theta_{k,\ell}$, $\hat{\theta}_{j,\ell}$, $\phi_{k,\ell}$, and $\hat{\phi}_{j,\ell}$ are independently drawn from the uniform distribution over $[0,\pi]$. The noise powers are identical across receivers, with $\sigma_k^2=\hat{\sigma}_j^2=\sigma^2$. The transmit signal-to-noise ratio (SNR) is set to $\frac{p}{\sigma^2}=10$ dB, the transmit region side length is $A=3\lambda$, $D=\frac{\lambda}{2}$, and $Q=10^3$, unless stated otherwise. We compare the proposed MA-enabled design with an FPA benchmark in which the BS employs a uniform linear array of $M$ fixed-position antennas with half-wavelength spacing. The transmit beamformer is initialized using maximal-ratio transmission, and the MA locations are initialized using the FPA baseline.

\begin{figure}[!t]
\centering
\includegraphics[height=0.19\textwidth]{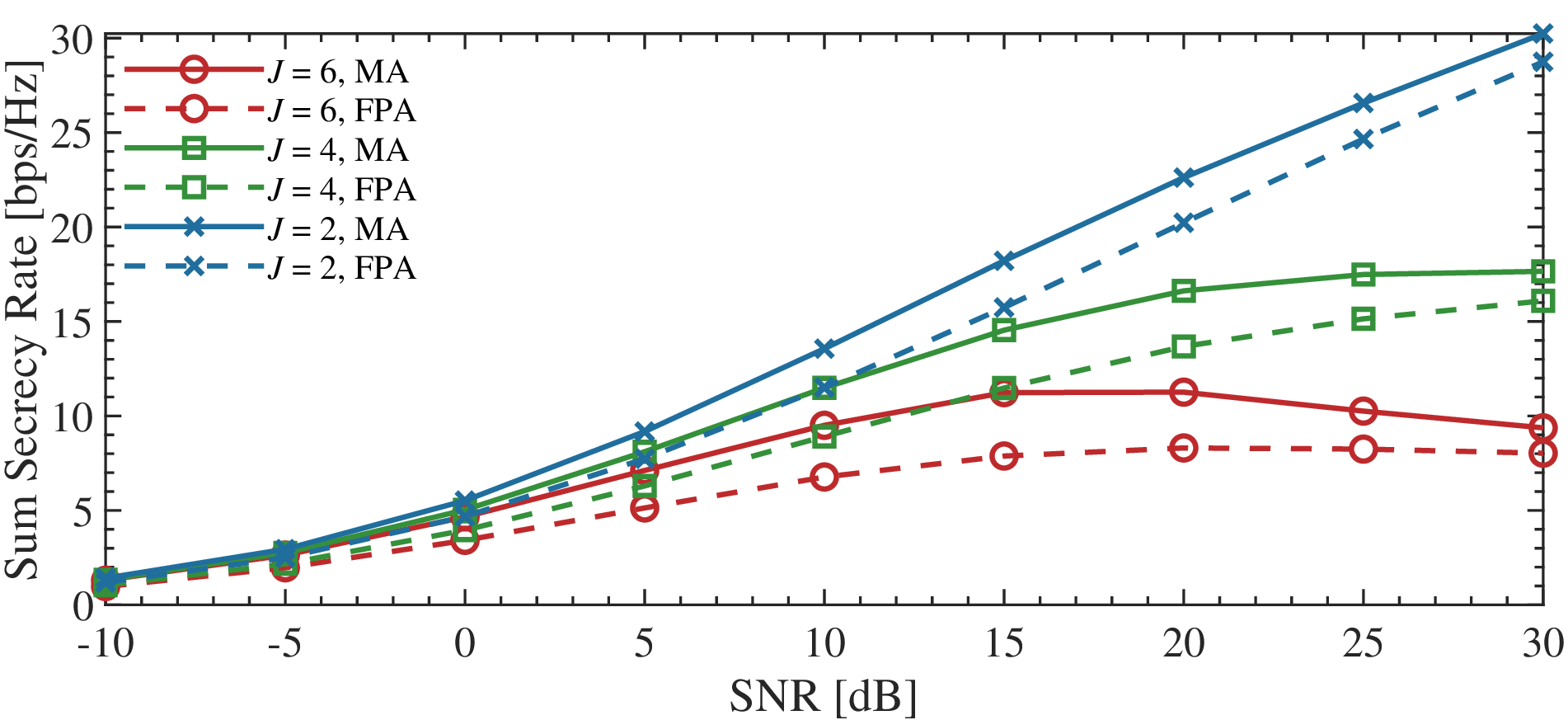}
\caption{Sum secrecy rate vs. the SNR. $K=6$ and $M=6$.}
\label{Figure3}
\vspace{-10pt}
\end{figure}

\begin{figure}[!t]
    \centering
    \subfigure[${\mathcal{R}}$ vs. $M$.]
    {
        \includegraphics[height=0.18\textwidth]{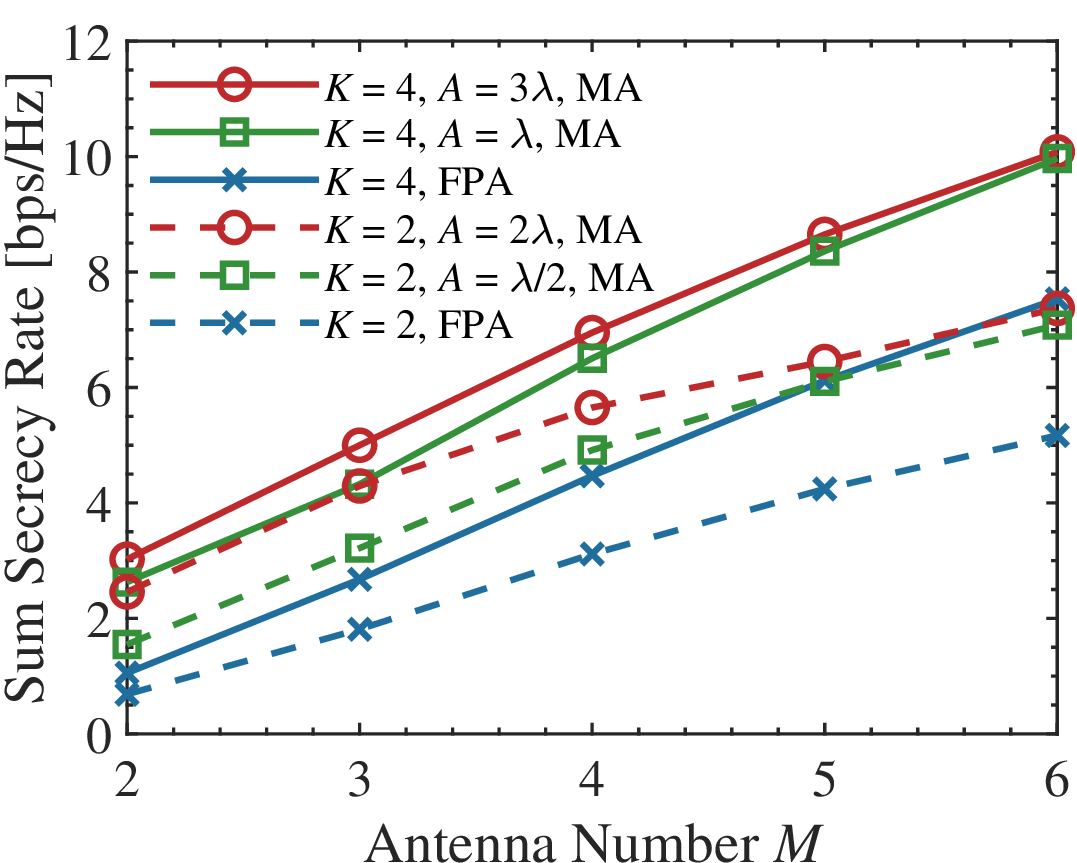}
	   \label{fig4a}	
    }
   \subfigure[${\mathcal{R}}$ vs. $K$.]
    {
        \includegraphics[height=0.18\textwidth]{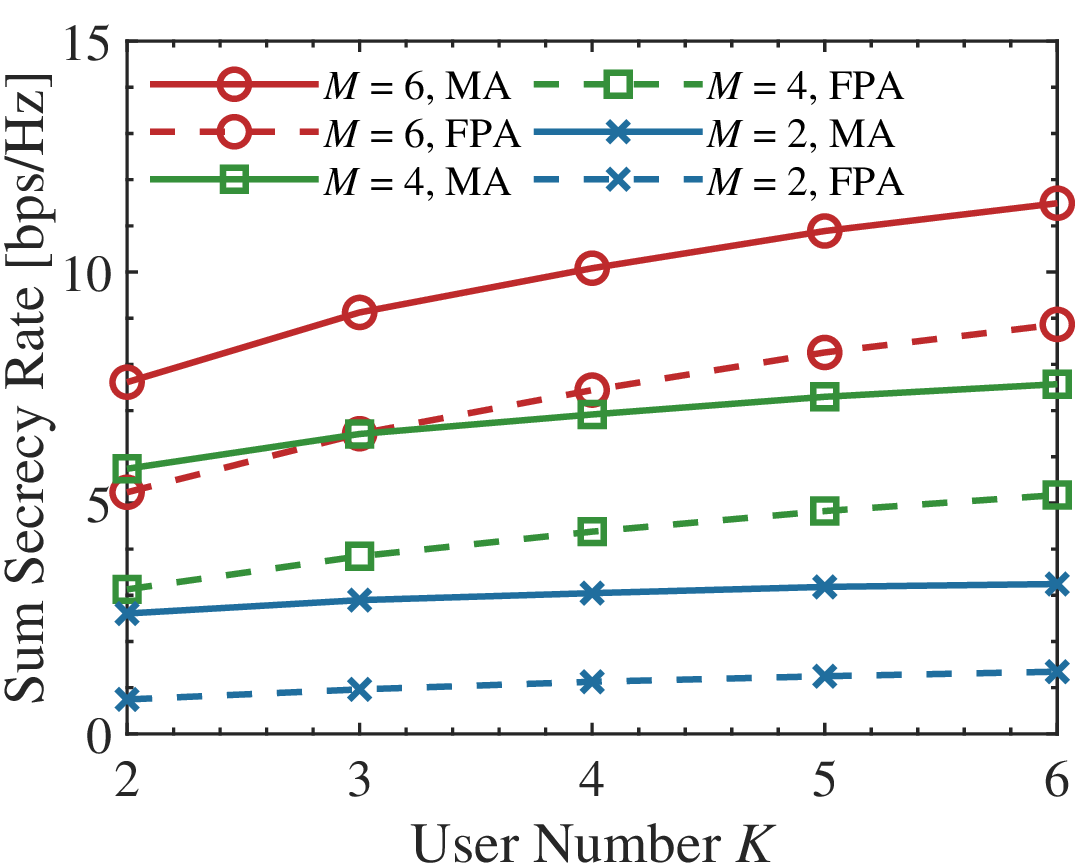}
	   \label{fig4b}	
    }
\caption{Sum secrecy rate vs. (a) the antenna number and (b) the UT number. $J=4$.}
    \label{figure4}
    \vspace{-10pt}
\end{figure}

\begin{figure}[!t]
    \centering
    \subfigure[Convergence.]
    {
        \includegraphics[height=0.18\textwidth]{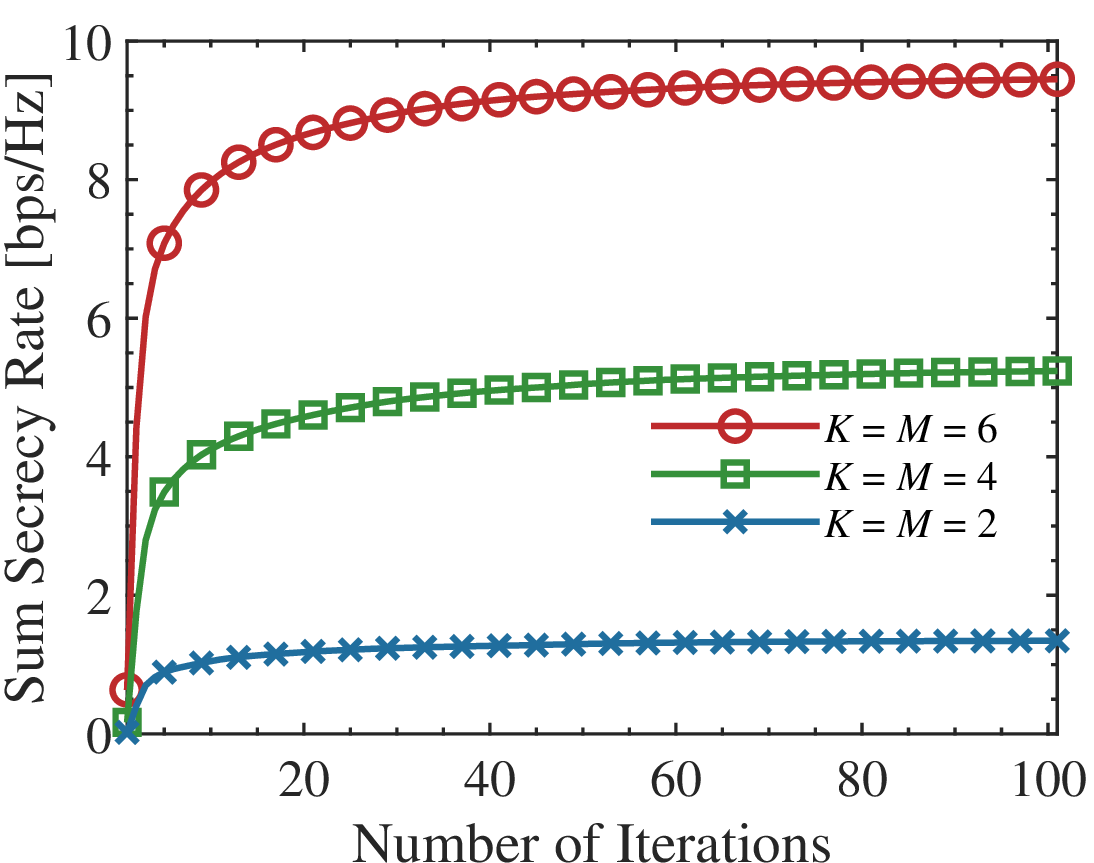}
	   \label{Figure5}	
    }
   \subfigure[${\mathcal{R}}$ vs. $Q$.]
    {
        \includegraphics[height=0.18\textwidth]{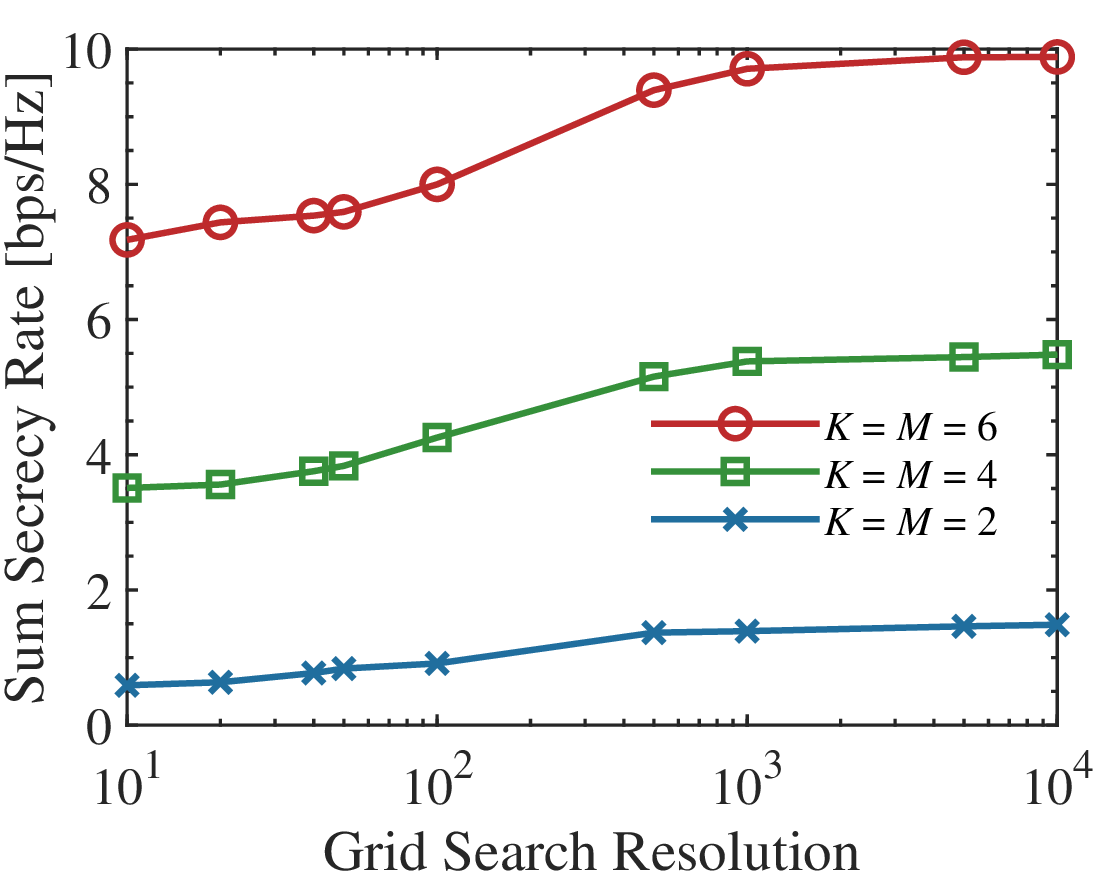}
	   \label{Figure6}	
    }
\caption{Sum secrecy rate vs. (a) the number of iterations and (b) the grid search precision $Q$. $J=6$.}
    \vspace{-10pt}
\end{figure}

{\figurename} {\ref{Figure3}} compares the sum secrecy rate of the proposed MA-enabled design and the conventional FPA benchmark versus the transmit SNR for different numbers of eavesdroppers. In all considered settings, the MA-enabled framework achieves a higher secrecy rate than the FPA scheme. This result confirms the advantage of MA deployment for multiuser secure transmission and also reflects the effectiveness of the proposed optimization algorithm. As the number of eavesdroppers increases, the secrecy rate decreases, since more eavesdroppers can collectively capture a larger portion of the leaked signal energy. {\figurename} {\ref{Figure3}} also shows a non-monotonic trend with respect to the transmit SNR when $J=6$. In this case, increasing the transmit power improves both the legitimate links and the eavesdropping links. At high SNR, the information leakage to the cooperating eavesdroppers becomes increasingly significant, which eventually reduces the achievable secrecy rate. In contrast, for smaller values of $J$, the leakage effect is less pronounced and the secrecy rate continues to increase within the considered SNR range.

{\figurename} {\ref{fig4a}} shows the sum secrecy rate versus the number of transmit antennas. The secrecy rate increases with $M$. {\figurename} {\ref{fig4a}} also indicates that a larger transmit region, characterized by a larger side length $A$, further improves the secrecy rate. This gain arises because more antennas and a larger MA placement region provide stronger spatial diversity and more degrees of freedom (DoFs) for leakage suppression. {\figurename} {\ref{fig4b}} depicts the sum secrecy rate as a function of the number of legitimate users. When the number of antennas is larger than the number of users, the available spatial DoFs are sufficient to suppress inter-user interference, and the secrecy rate increases with $K$. In contrast, in the RF-limited case with $M=2$, the secrecy rate decreases as $K$ grows, since the system cannot simultaneously manage multiuser interference and information leakage. In all cases, the MA-enabled design consistently outperforms the FPA benchmark in terms of the secrecy rate. 

{\figurename} {\ref{Figure5}} shows the convergence behavior of the proposed BCD-based algorithm. As the number of iterations increases, the secrecy rate monotonically improves and eventually converges. {\figurename} {\ref{Figure6}} shows that increasing the grid search resolution $Q$ improves the secrecy rate, while the marginal performance gain diminishes for large $Q$.

\section{Conclusion}
We proposed an MA-enabled multiuser secure transmission framework that jointly optimizes the digital beamforming and the MA locations to maximize the sum secrecy rate. Numerical results show that the proposed design achieves clear secrecy-rate gains over conventional FPAs. Future work may consider robust designs under imperfect CSI, practical MA mobility constraints, non-cooperative eavesdropping scenarios, and wideband or frequency-selective channels.

\bibliographystyle{IEEEtran}
\bibliography{mybib}
\end{document}